\documentclass[twocolumn, tighten]{aastex62}

\usepackage{verbatim}
\usepackage{color}
\usepackage{natbib}
\usepackage{shortbold}
\usepackage{booktabs}

\begin{document}

\title{A Machine Learning Dataset Prepared From the NASA Solar Dynamics Observatory Mission}

\correspondingauthor{Richard Galvez}
\email{richard.galvez@nyu.edu}

\author[0000-0002-4780-9566]{Richard Galvez*}
\affil{Center for Data Science, New York University}

\author[0000-0001-5028-5161]{David F. Fouhey*}
\affiliation{University of Michigan}

\author[0000-0002-9672-3873]{Meng Jin}
\affiliation{Lockheed Martin Solar \& Astrophysics Laboratory}
\affiliation{SETI Institute}

\author[0000-0002-4829-5739]{Alexandre Szenicer}
\affiliation{University of Oxford}

\author[0000-0002-4716-0840]{Andr\'es Mu\~noz-Jaramillo}
\affiliation{Southwest Research Institute}

\author[0000-0003-2110-9753]{Mark C. M. Cheung}
\affiliation{Lockheed Martin Solar \& Astrophysics Laboratory}
\affiliation{Hansen Experimental Physics Laboratory, Stanford University}

\author[0000-0001-9021-611X]{Paul J. Wright}
\affiliation{SUPA School of Physics \& Astronomy, University of Glasgow}

\author[0000-0002-5662-9604]{Monica G. Bobra}
\affiliation{Hansen Experimental Physics Laboratory, Stanford University}

\author[0000-0002-0671-689X]{Yang Liu}
\affiliation{Hansen Experimental Physics Laboratory, Stanford University}

\author[0000-0002-3783-5509]{James Mason}
\affiliation{NASA Goddard Space Flight Center}

\author[0000-0002-5362-4816]{Rajat Thomas}
\affiliation{University of Amsterdam}


\begin{abstract}
    In this paper we present a curated dataset from the NASA Solar Dynamics Observatory (SDO) mission in a format suitable for machine learning research. Beginning from level 1 scientific products we have processed various instrumental corrections, downsampled to manageable spatial and temporal resolutions, and synchronized observations spatially and temporally. We illustrate the use of this dataset with two example applications: forecasting future EVE irradiance from present EVE irradiance and translating HMI observations into AIA observations. For each application we provide metrics and baselines for future model comparison. We anticipate this curated dataset will facilitate machine learning research in heliophysics and the physical sciences generally, increasing the scientific return of the SDO mission. This work is a direct result of the 2018 NASA Frontier Development Laboratory Program. Please see the appendix for access to the dataset.
\end{abstract}

\section{Introduction}

Launched in 2010, NASA's Solar Dynamics Observatory~\citep[SDO;][]{SDO} has been continuously monitoring the Sun's activity and delivering valuable scientific data for heliophysics researchers with the use of three instruments: \begin{itemize}
    \item The Atmospheric Imaging Assembly~\citep[AIA][]{AIA} which captures $4096\times4096$ resolution images (with $0.6$ arcsec pixel size) of the full Sun in two ultraviolet (UV; centered at $1600~\&~1700$ \AA) wavelength bands, seven Extreme Ultraviolet (EUV) wavelength bands (centered at 94, 131, 171, 193, 211, 304 and 335 \AA) and one visible wavelength (centered at $4500$ \AA).
    \item The Helioseismic and Magnetic Imager~\citep[HMI][]{HMI} captures visible wavelength filtergrams of the full Sun at $4096\times4096$ resolution (pixel size of $0.5$ arcsec), which are then processed into a number of products, including photospheric dopplergrams, line-of-sight magnetograms and vector magnetograms~\citep{Hoeksema:2014}.
    \item The EUV Variability Experiment~\citep[EVE;][]{EVE} monitors the solar EUV spectral irradiance from $1$ to $1050$  \AA. This is done by utilizing multiple EUV Grating Spectrographs (MEGS) which disperse EUV light from the full disk of the Sun and its corona onto a 1024 x 2048 CCD.
\end{itemize}  

Calibrated level 1 scientific data from the AIA and HMI instruments are accessible from the Joint Science Operations Center\footnote{\url{http://jsoc.stanford.edu}} (JSOC) at Stanford University, Lockheed Martin Solar \& Astrophysics Laboratory, and affiliate science data centers; while science data from the EVE instrument are accessible from the EVE Science Operations Center\footnote{\url{http://lasp.colorado.edu/home/eve/data}} at the Laboratory for Atmospheric and Space Physics (LASP) at the University of Colorado, Boulder.

The SDO mission has been scientifically prolific. In the eight years after launch, over 3000 refereed scientific publications\footnote{\url{https://sdo.gsfc.nasa.gov/mission/publications.php}} have made use of SDO data. This success can be attributed to the reliability of the spacecraft and its instruments, the consistency and quality of the observations, the mission's open data policy, and the ease of online data access from the affiliate science data centers. The large volume of structured, calibrated scientific data (over 12 Petabytes and counting) is poised for exploratory analysis from machine learning methods as well as more traditional approaches. In early pioneering works, supervised learning techniques have been applied to the prediction of solar flares using HMI vector magnetograms~\citep[e.g.,][]{BobraCouvidat:2015}, as well as HMI and AIA imagery in ~\cite{Jonas:2018}. Deep learning applications have began to emerge from the heliophysics community as well, with exemplary cases illustrated in \cite{Colak2013} and \cite{Huang_2018}, and \cite{DeepEM} presenting a more recent treatment using the dataset presented here.

While level 1 data are easily accessible, pre-processing these data for scientific analysis often requires specialized heliophysics knowledge. The necessity for such pre-processing may act as an unnecessary hurdle for non-heliophysics machine learning researchers whom may wish to experiment with datasets from the physical sciences, but are unaware of domain-specific nuances (e.g., that images must be spatially and temporally adjusted). 

The first contribution of this paper is a curated SDO dataset that is mission-ready for machine learning applications. Our aim is to supply this standardized dataset for heliophysicists who wish to use machine learning in their own research, as well as machine learning researchers who wish to develop models specialized for the physical sciences. In Section \ref{sec: data exploration}, we examine current available data products, the pitfalls for their direct use in machine learning tasks, as well as what corrections and adjustments they warrant. These corrections are incorporated into our data preparation procedures that are discussed in Section \ref{sec: data prep}. 

The second contribution of this paper are protocols, metrics, and baseline models. We introduce evaluation protocols and metrics in Section \ref{sec: methods}, and baseline models in Section \ref{sec: results} where we tackle the tasks of predicting irradiance using present and future EVE data, as well as translating 3 HMI channels into 9 AIA channels. We believe these models contain generic enough components with providing useful benchmarks, and highlighting the most dangerous pitfalls, for most subsequent SDO machine learning applications.

By providing these standardized data products along with accompanying protocols, metrics, and baselines, our aim is to remove unnecessary hurdles for future machine learning research in heliophysics, and the physical sciences more broadly.

\begin{figure}
\vspace*{-10pt}
\hspace*{-20pt}
\includegraphics[width=10cm]{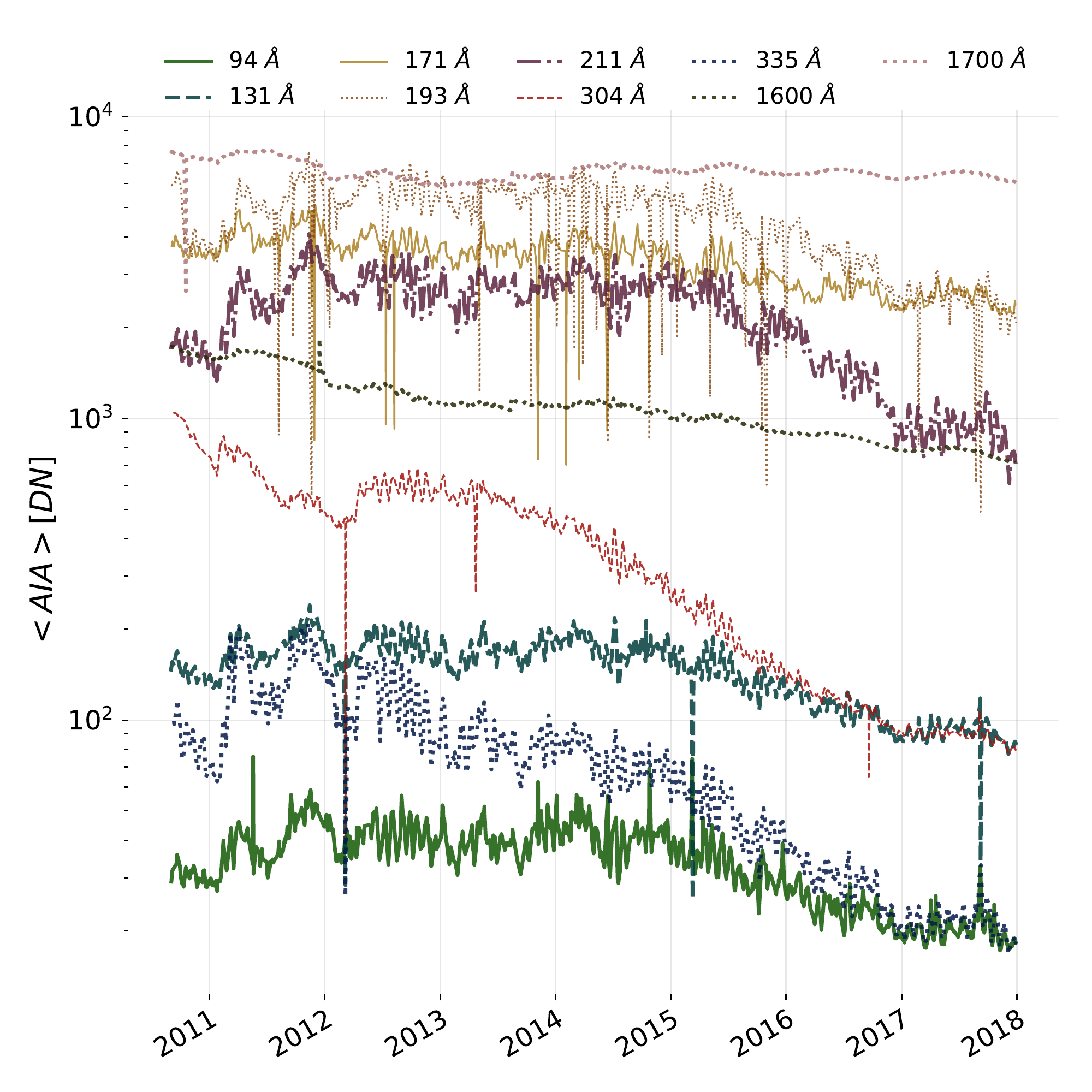}
    \caption{The average AIA wavelength band data counts for level 1 data products over time, presented here randomly downsampled to 1k observations. The secular downward trend is caused by instrument degradation over time, while the spurious per-channel drops are caused by the instrument's automatic exposure control mechanism.   
    \label{fig:corrupt_aia_means}}
\end{figure}

\section{Examination of Raw Data Products}\label{sec: data exploration}

\begin{figure}[t]
\centering
\includegraphics[width=\columnwidth]{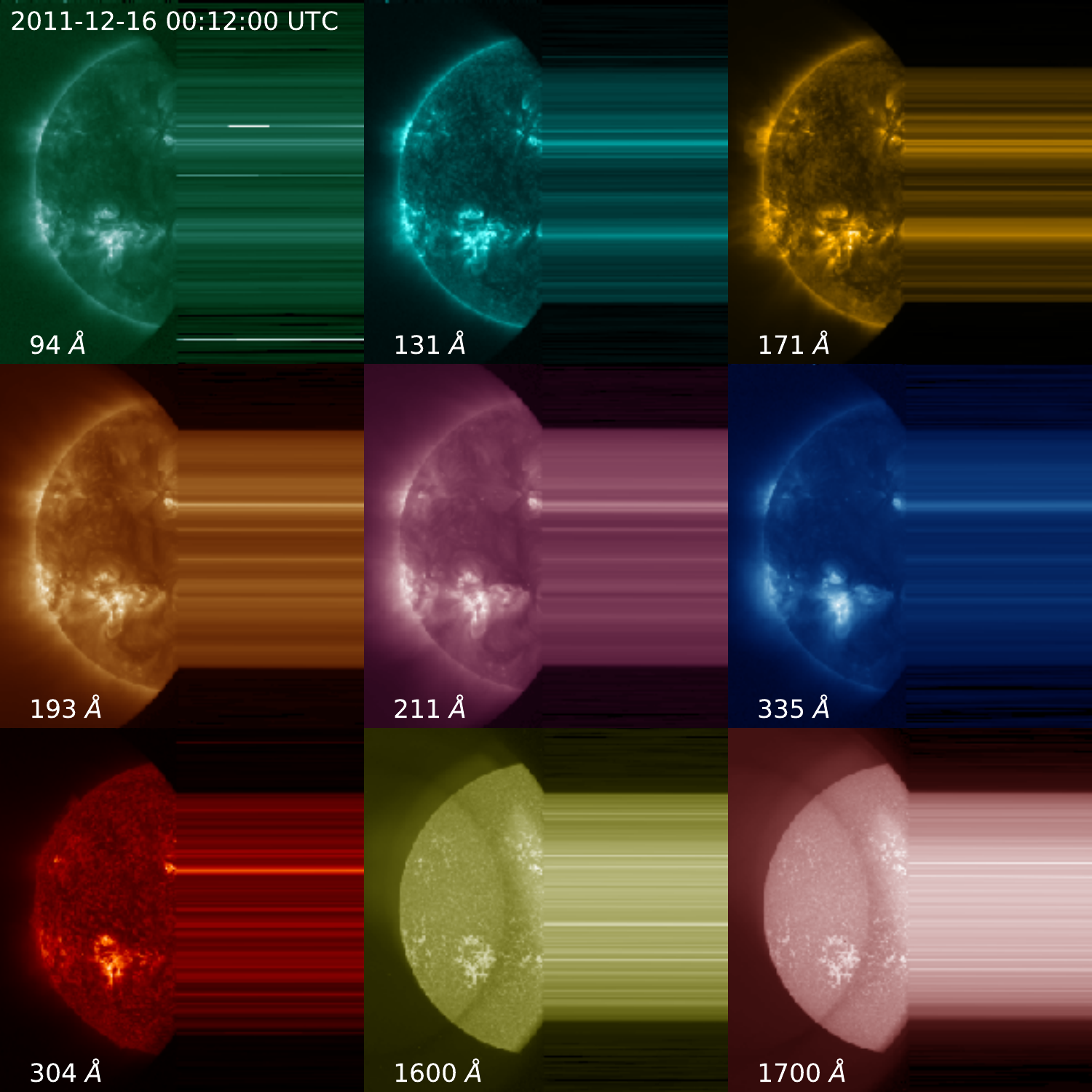}
    \caption{Example of a corrupt observation from the AIA instrument. Utilizing such observations during the training phase of a machine learning model may compromise its predictive capability.
    \label{fig:aia_example}}
\end{figure}

We first examine existing raw data products available from SDO for each of the three instruments (level 1 science data products for AIA; \texttt{hmi.B\_720s} for HMI; and the EVE version 5 IDL saveset). 

While heliophysics researchers are likely aware of corrections that must be applied to this data, and the fact that AIA measurements have heterogenous exposure times, it is unrealistic to expect the same from researchers in other fields ~\citep[e.g. the data set of][were compiled from quicklook JPEG2000 images that have compressed dynamic range and do not account for instrumental degradation]{Kucuk:2017}. We therefore process these corrections by identifying and removing corrupt observations (e.g., images taken during instrument anomalies), adjust detected intensities for heterogeneous exposure times, and fix instrument artifacts that introduce spurious trends. 

If such corrupt observations or various sources of heterogeneity are not removed, any subsequent machine learning model will likely learn to emulate these incorrect observations as well as any spurious trends, and will not be able to isolate the physical dynamics. Exposing such corrupt data during model training may also compromise predictive quality; or worse, the model may even learn to emulate non-physical aberrations and instrumental noise. See Figure \ref{fig:aia_example} for an example of one such unwanted AIA observation.

To identify each instrument's possible issues, we visualize each instrument's data by taking the average channel values (i.e., AIA wavelength 
band data counts, HMI vector field components, and EVE irradiance values) and plot them over time. We then identify the underlying causes of nonphysical aberrations and what necessary corrections are needed to standardize the data. Below we report our analysis for AIA and outline where HMI required similar adjustments; EVE level-3 data products already address all instrumental issues so we only adjusted for time synchronicity. We describe these corrections in Section \ref{sec: data prep}.

The average channel values for the AIA level 1 data products, as plotted in Figure~\ref{fig:corrupt_aia_means}, shows the data heterogeneity as well as the presence of corrupt observations. These are visible in this figure as isolated downward spikes, while the secular downward trend is indicative of degradation over the lifetime of the instrument. 

Corrupt observations arise due to a variety of reasons such as data reported during calibration maneuvers, eclipse periods, or the occasional instrument anomaly. Such data, flagged by a non-zero value of the {\tt QUALITY} keyword for both the AIA and HMI instruments, are not intended for scientific analysis and are removed from our dataset. 
One of the main sources of heterogeneity in AIA data responds to its instrument design: AIA instrumentation is not designed to directly measure irradiance, but rather data numbers (DNs) as tabulated by the activation of the CCD instrument. While it is intended that DN values are proportional to the flux of photons at a specific wavelength~\citep{Boerner:2012}, the factor of proportionality is not constant in time. For instance, the camera exposure time $t_{\rm exp}$ is not constant due to instrumental automatic exposure control (AEC); e.g.in times of flares when certain regions on the Sun become especially bright, AEC reduces the nominal exposure time from a few seconds to tens of milliseconds. Due to these factors, when the AEC is activated, the mean registered DNs are drastically reduced; an effect easily compensated for by adjusting for the exposure time.

The visible downward trend in Figure~\ref{fig:corrupt_aia_means} is caused by the gradual in-orbit degradation of the AIA instrument. This degradation is purely due to CCD corrosion over time. Because AIA is calibrated against EVE, which is itself bootstrapped to a program of regular EUV spectral irradiances as measured from sounding rocket launches ~\citep{Boerner:2014}, the time-dependent profile of the AIA instrument is well understood independently of the solar-cycle. This instrumental understanding allows us to correct for AIA instrument degradation by simply applying the \texttt{aia\_get\_response} routine in the \texttt{SolarSoft} software package \citep{Freeland1998}. 

Lastly, there is a more subtle non-monotonic heterogeneity caused by SDO's orbit around the Sun. SDO is in a geosynchronous orbit around the Earth, which itself is in a slightly elliptical orbit around the solar system barycenter. The elliptical orbit causes the size of the Sun (in DNs registered on the CCD) to vary by about 10\% over the course of a year. This is not an intrinsic feature of solar evolution. We compensate for this effect by resizing AIA and HMI images such that the size of the solar disk is fixed to some size $R_s$. In particular, we can scale the AIA and HMI images by a factor of $R_s/R_{\textrm{obs}}$, where $R_{\textrm{obs}}$ can be obtained from \texttt{RSUN\char`_OBS} keyword in the level 1 FITS header.

\section{Processed Data Preparation}\label{sec: data prep}

We now describe how our processed dataset is produced in detail. First, we describe how the needed corrections outlined in Section \ref{sec: data exploration} are applied to each instrument, and how temporal synchronization is computed. We first begin by removing the non-zero {\tt QUALITY} observations from both AIA and HMI. We then spatially downsample to produce a more manageable dataset, while being careful to emulate what a lower resolution instruments would observe. 

\subsection{AIA}

We begin with the $4096\times4096$ pixel level 1~\citep[][dark subtracted, flat fielded and despiked]{AIA} data products and process them as described below:

\begin{figure*}[t]
\vspace*{-10pt}
\hspace*{-25pt}
\includegraphics[width=19cm]{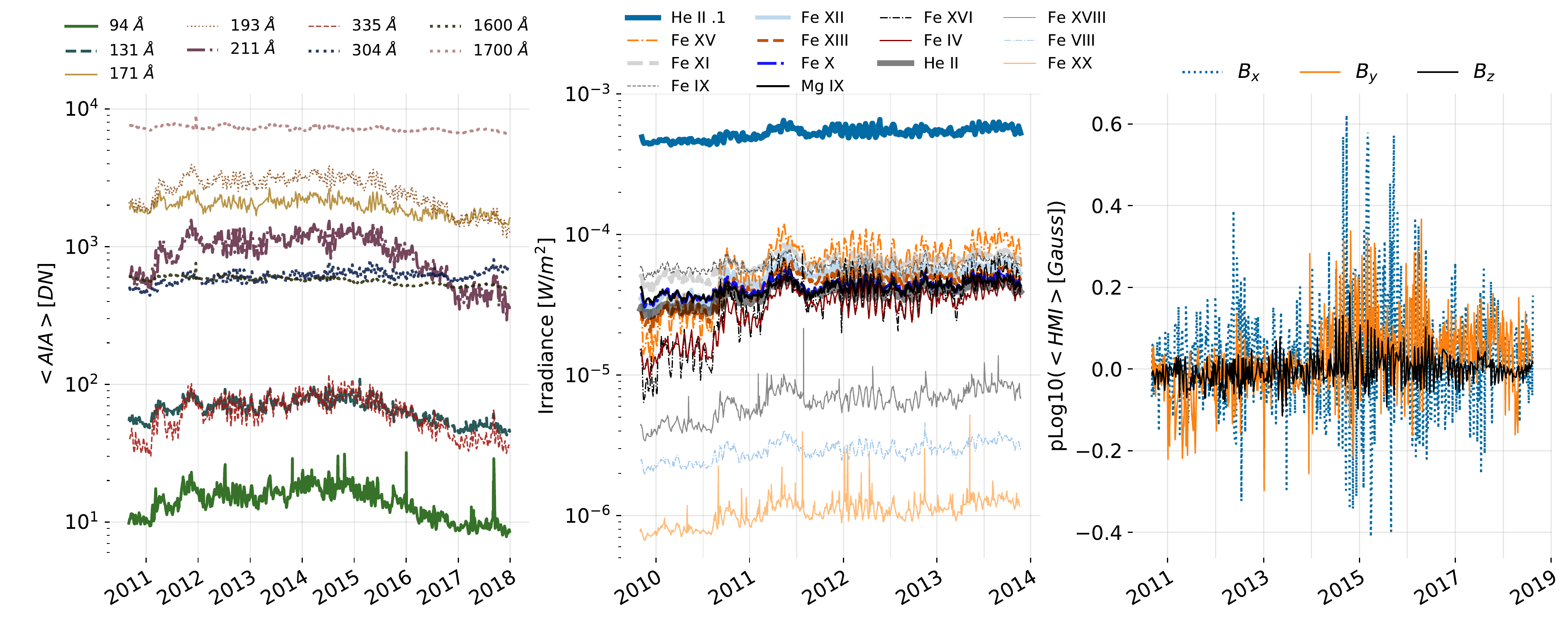}
    \caption{Mean intensity variation after correcting for exposure time, degradation, and orbital variation for AIA (Left Panel) and EVE (Middle Panel).  Middle Panel:  We display only the 14 emission lines covered by the MEGS-A instrument, for illustration purposes, out of the 37 in our data product (see Table \ref{tab:EVEWave}). Right Panel: The signed pseudo logarithm of the mean field values for $B_x$, $B_y$, and $B_z$ from HMI after correcting for orbital variation.
    \label{fig:good_aia_mean}}
\end{figure*}

\begin{enumerate}

\item The raw images are rotated and resized onto a common grid (still $4096\times4096$ pixels) such that the pixel size is $0.600$ arcsec, and $\hat{x}$ and $\hat{y}$ (the first and second image dimensions) are aligned with the solar west and north directions, respectively.

\item Images are re-binned by averaging neighboring $4\times4$ pixel blocks such that the resultant image has size $1024\times1024$ pixels (with a final pixel size of $2.400$ arcsec). Resultant images are processed at a $2$ min cadence producing the so-called Synoptic series\footnote{\url{http://jsoc.stanford.edu/data/aia/synoptic/}}.

\item The AIA images are then normalized by exposure time and corrected for instrument degradation, while corrections for elliptical orbital variation are applied with a fixed disk size of $R_s$ of 976 arcsec.

\item Finally, the images are downsampled again by \textit{summing} in local blocks, which emulates the expected observation of a lower resolution instrument. The final interpolated images have $512\times512$ pixels with pixel size of $\sim$4.8 arcsec.
\end{enumerate}

\subsection{HMI}

We start with the original HMI JSOC data series \texttt{him.B\_720s}, which provides the magnetic vector field strength, inclination angle, and azimuth \citep{Hoeksema:2014}. We process this to calculate full-disk vector field observations in $B_x$, $B_y$, and $B_z$ components with 12 minute cadence. The $+x$ direction points to the solar west, $+y$ to the north, and $+z$ out of the image plane (i.e., line-of-sight). Additionally, as with the AIA instrument, although the original image size is 4096$\times$4096, the pixel resolution is different. We therefore further co-aligned HMI and AIA data so that they have the same spatial sampling. The major processing steps for the HMI observations are as follows:

\begin{enumerate}

\item We begin by converting the original HMI JSOC data series \texttt{hmi.B\_720s} vector field data with the disambiguation solution of \texttt{disambig.fits} to $B_x$, $B_y$, and $B_z$ components, spatially co-aligning with AIA observations using the \texttt{FITS} header information.

\item The HMI images were also corrected for orbital variation with a fixed disk size $R_s$ of 976 arcseconds throughout.

\item Finally, we downsampled the data by \textit{averaging} in local blocks, which emulates the expected observation at the target lower resolution. The final interpolated images have $512\times512$ pixels with pixel size of $\sim$4.8 arcsec.
\end{enumerate}

\begin{table}[t]
    \caption{EVE Emission lines, their wavelength, and temperature of the emission plasma.  The MEGS-A emission lines used in Section \ref{sec:EVEPRed} are highlighted in purple. \label{tab:EVEWave}}
    \centering
    \begin{tabular}{
    @{}l@{~~}c@{~~}c@{~~}} 
            Emission Line & Wavelength & Temperature\\
            \midrule
    \textcolor{purple}{\ion{Fe}{18}} & 93.9 \AA & 6.46$\times$10$^6$ K\\
    \ion{Ne}{7} & 127.7 \AA & 5.01$\times$10$^5$ K\\
    \textcolor{purple}{\ion{Fe}{8}}  & 131.2 \AA & 3.71$\times$10$^5$ K\\
    \textcolor{purple}{\ion{Fe}{20}} & 132.8 \AA & 9.33$\times$10$^6$ K\\
    \ion{Ne}{5} & 148.7 \AA & 3.16$\times$10$^5$ K\\
    \textcolor{purple}{\ion{Fe}{9}}  & 171.0 \AA & 6.46$\times$10$^5$ K\\
    \textcolor{purple}{\ion{Fe}{10}} & 177.2 \AA & 9.77$\times$10$^5$ K\\
    \textcolor{purple}{\ion{Fe}{11}} & 180.4 \AA & 1.17$\times$10$^6$ K\\
    \textcolor{purple}{\ion{Fe}{12}} & 195.1 \AA & 1.35$\times$10$^6$ K\\
    \textcolor{purple}{\ion{Fe}{13}} & 202.0 \AA & 1.55$\times$10$^6$ K\\
    \textcolor{purple}{\ion{Fe}{14}} & 211.3 \AA & 1.86$\times$10$^6$ K\\
    \textcolor{purple}{\ion{He}{2}}  & 256.3 \AA & 5.62$\times$10$^4$ K\\
    \textcolor{purple}{\ion{Fe}{15}} & 284.2 \AA & 5.01$\times$10$^4$ K\\
    \textcolor{purple}{\ion{He}{2}}  & 303.8 \AA & 1.99$\times$10$^6$ K\\
    \textcolor{purple}{\ion{Fe}{16}} & 335.4 \AA & 2.69$\times$10$^6$ K\\
    \ion{Fe}{16} & 360.8 \AA & 2.69$\times$10$^6$ K\\
    \textcolor{purple}{\ion{Mg}{9}}  & 368.1 \AA & 9.77$\times$10$^5$ K\\
    \ion{Mg}{9} & 443.7 \AA & 1.00$\times$10$^6$ K\\
    \ion{Ne}{7} & 465.2 \AA & 3.98$\times$10$^5$ K\\
    \ion{Si}{12} & 499.4 \AA & 1.99$\times$10$^6$ K\\
    \ion{O}{3} & 525.8 \AA & 7.94$\times$10$^4$ K\\
    \ion{O}{4} & 554.4 \AA & 1.99$\times$10$^5$ K\\
    \ion{He}{1} & 584.0 \AA & 1.99$\times$10$^4$ K\\
    \ion{O}{3} & 599.6 \AA & 7.94$\times$10$^4$ K\\
    \ion{Mg}{10} & 624.9 \AA & 1.26$\times$10$^6$ K\\
    \ion{O}{5} & 629.7 \AA & 2.51$\times$10$^5$ K\\
    \ion{O}{2} & 718.5 \AA & 6.31$\times$10$^4$ K\\
    \ion{N}{4} & 765.1 \AA & 1.58$\times$10$^5$ K\\
    \ion{Ne}{8} & 770.4 \AA & 6.31$\times$10$^5$ K\\
    \ion{O}{4} & 790.2 \AA & 1.99$\times$10$^5$ K\\
    \ion{H}{1} & 972.5 \AA & 5.01$\times$10$^4$ K\\
    \ion{C}{3} & 977.0 \AA & 5.01$\times$10$^4$ K\\
    \ion{H}{1} & 1025.7 \AA & 5.01$\times$10$^4$ K\\
    \ion{O}{6} & 1031.9 \AA & 2.51$\times$10$^5$ K
    \end{tabular}
\end{table}

\newpage
\subsection{EVE}

\begin{figure}[t]
\hspace{-18pt}
\includegraphics[width=9cm]{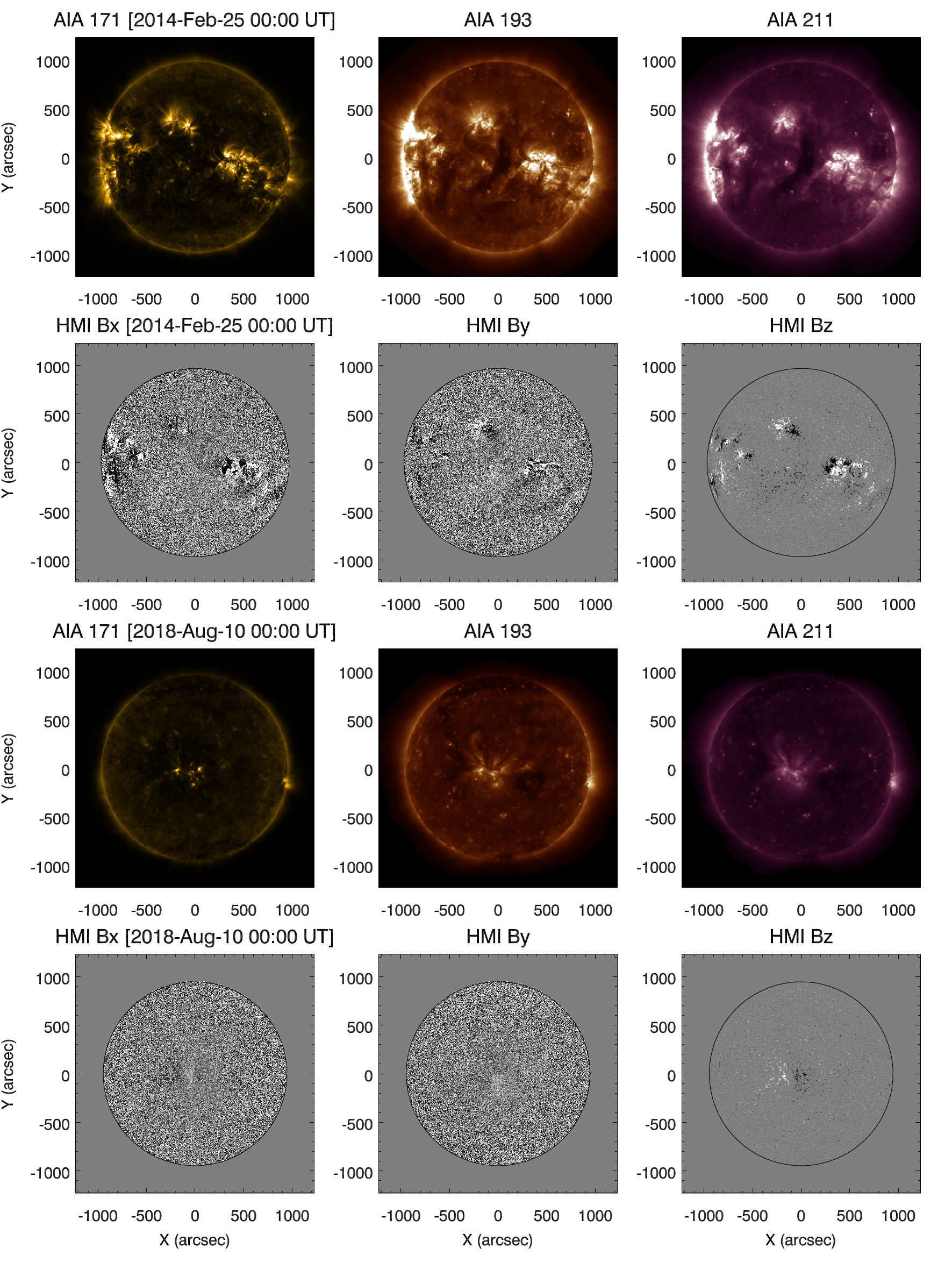}
\caption{Upper Panel: Co-aligned AIA and HMI dataset around solar maximum (2014-February-25 00:00 UT) of solar cycle 24. Three selected AIA wavebands (171, 193, 211 \AA) are shown; Bottom Panel: Co-aligned AIA and HMI dataset around solar minimum (2018-August-10 00:00 UT) of solar cycle 24. The black cycle in the HMI magnetogram shows the location of solar limb.}
\label{fig: hmiaia}
\end{figure}

EVE spectra are assembled form a battery of instruments including the Multiple EUV Grating Spectrographs (MEGS-A, B, P), Solar Aspect Monitor (SAM), and the Extreme ultraviolet SpectroPhotometer (ESP).  Each of these instruments covers a different wavelength range in the EUV spectra and are cross calibrated to produce EVE's data products.

The EVE data released in this dataset is extracted from a specially prepared EVE version 5 IDL saveset, including 39 emission lines (see Table \ref{tab:EVEWave}) during the time window between 2010-05-01 and 2014-05-26.  The end date of this dataset corresponds to the failure of the MEGS-A instrument, which covered the range between 30$\AA$ and 370$\AA$. The EVE data have already been calibrated with physical units of $W m^{-2}$, scaled to 1 AU, and corrected for degradation, requiring no subsequent calibration. The only processing we perform is to convert from IDL to NumPy Arrays and temporally synchronize with the AIA and HMI observations.

\subsection{Temporal Downsampling and Synchronization}

One of the goals of this paper is to produce a dataset that is temporally and spatially synchronized for the three SDO data products at manageable resolutions. While our scaling to a fixed solar disk size automatically ensures the spatial synchronization of AIA and HMI, all SDO data instruments observe at different cadences (AIA: 2 minutes, HMI: 12 minutes, EVE: 10 seconds) and are not necessarily aligned in time.

In order to perform the temporal synchronization, we downsample AIA to a 6 minute cadence and match the nearest EVE observation within a mean/max time window of 8.5s/12s. This yields a final dataset consisting of AIA and EVE observations each at 6 minute cadence, with accompanying HMI observations occurring at every other time step\footnote{With the exception of non-zero {\tt QUALITY} observations for either AIA or HMI.} with a 12 minute cadence.

\subsection{Data} 

This produces the final dataset for the paper totalling $\sim$6.5 TB, made available through the Stanford Digital Repository\footnote{\url{https://purl.stanford.edu/nk828sc2920}} (please see appendix for a list of URLs). The data are individually packed monthly and for each waveband/component of AIA/HMI, all in the {\texttt NumPy} format. The EVE data are packed in a single {\tt TAR} file. We show the average value as a function of time for the three products in Figure~\ref{fig:good_aia_mean}, which demonstrates the removal of spurious trends and artifacts. We also show in Figure~\ref{fig: hmiaia} an example of co-aligned AIA and HMI observations. The upper panel shows the observation near the solar maximum of cycle 24 (2014 February 25 00:00 UT), exhibiting several active regions with strong magnetic field magnitudes and associated EUV emission. The bottom panel shows the observation near the solar minimum of cycle 24 (2018 August 10 00:00 UT), displaying only one active region with a comparatively weak magnetic field and EUV brightness. 

\section{Protocols and Metrics}\label{sec: methods}

We expect that this data will be of interest for machine learning applications in heliophysics and a simple-access dataset for the testing of machine learning models in the physical sciences. To facilitate this, we have defined standard protocols and metrics to aid future work with this data.

~~ ~~
\\
\noindent {\bf Data splits:} There is large temporal coherence in the data since large-scale structures on the Sun evolve at timescales beyond days and months. This leads to issues with randomly sampled splits of the data, often done in machine learning settings with uncorrelated data. In particular, randomly sampled splits will lead to training and testing observations that are separated by days or even minutes. While these observations are indeed distinct points in time, they are generated by virtually the same large-scale structures. 

In practice, this means that experiments on randomly split data will be unable to identify overfitting and likely lead to overly optimistic estimates of generalization performance. The specific issue is that when deploying a model, one tests it on large-scale structures and conditions that are different than the training data. However, if the data is split randomly, the model is never actually evaluated on unseen large-scale structures due to temporal coherence. Therefore, there is no indication of whether the model's performance is due to generalizing well or if it is simply explained by the model overfitting to the particularities of the limited large-scale structures observed at training time.

To preclude this, we have split our data in temporal blocks that break this correlation, consisting of (i) a training set used to fit model parameters (e.g., the filter weights of a convolutional neural network); (ii) a validation set used to set model hyperparameters (e.g., the learning rate for training a network); and (iii) a test set used to evaluate out-of-sample model performance.

All of our data splits are performed over the years (2011-2014), the time period for which all three SDO instruments (AIA, HMI, EVE) were active. This time period provides a dataset large enough to support the training of modern models that require copious amounts of data. We set aside years 2012 and 2013 for testing purposes, supplying a wide variety of solar conditions. Years 2011 and 2014 are split into training and validation such that 70\% of available EVE observations are used for training (until mid-December 2011) and 30\% are used for validation. Of course, other splits are possible, especially for problems not relying on EVE observations. We therefore encourage the community to experiment with various data splits, with the cautionary advice that splits should be constructed in temporal blocks as opposed to random sampling.

\begin{figure*}[]
\hspace*{-15pt}
\includegraphics[width=18.7cm]{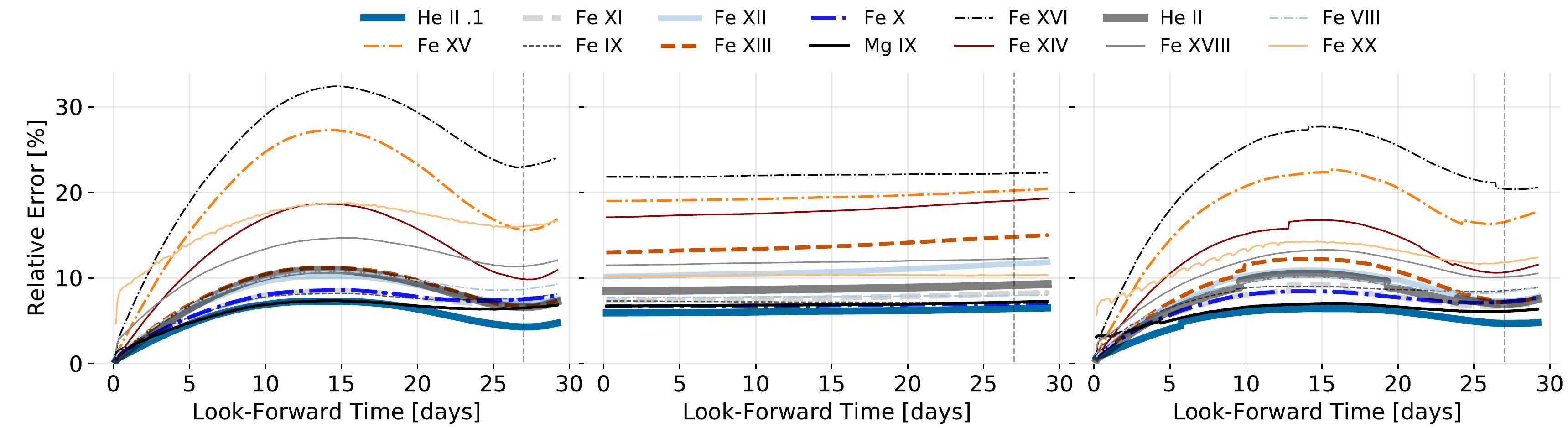}
    \caption{Results from the 2 hour EVE Prediction experiment. Left Panel: Persistence model, Middle Panel: Average assumption model, Right Panel: Ridge regression model.  This prediction exercise is performed for illustration purposes on 14 MEGS-A emission lines (see Table \ref{tab:EVEWave}). The forecast errors of all intervals, N hours apart, contained in the years 2012 and 2013 are averaged to produce the average error plotted for an N hour look-forward time in these figures.  The average model surprisingly predicts 7 of the 14 EVE lines within 10\% error not showing much overall variation, while the persistence model achieves this same performance for 10 lines. The ridge regression model often outperforms the persistence model overall, but not in all conditions and not by a substantial margin. The linear and persistence models both show periodic trends consistent with one solar rotation. See Section \ref{sec:EVEPRed} for discussion. 
    \label{fig:future_predictions}}
\end{figure*}

~~ ~~
\\
\noindent {\bf Metrics:} All of the metrics reported in Section \ref{sec: results} are derived from the normalized absolute error, or $|y_i-\hat{y_i}|/y_i$ where $\hat{y}$ is the model prediction and $y$ the measurement for data point $i$. For scalar quantities like EVE prediction, that are intrinsically already averaged over the Sun, we report the average normalized absolute error over all samples in the test set. 

For images (e.g., AIA prediction) that are not already spatially integrated, we report a number of metrics. 
First, we report the average normalized absolute error averaging first over each predicted image's valid pixels, and then over the images. In computer vision research, this average has been noted to often poorly characterize the performance on most pixels \cite{Scharstein02} since it can be arbitrarily changed by a small number of large errors. Thus,  we also report the percentage of good pixels metric, or the fraction of image pixels with normalized absolute error less than a fixed percentage $t$ for $t = 10\%, 20\%, 50\%$.

\newpage
\section{Results}\label{sec: results}

In this section, we provide baseline metrics for simple machine learning applications utilizing the proposed dataset, all implemented in the {\tt PyTorch} library \cite{paszke2017automatic}. These examples were chosen to illustrate what performance metrics should be expected from future models, as well as supplying simple examples for typical use cases. To this end, we have selected and evaluated two problems that demonstrate the temporal nature of the data as well as the alignment between the two spatially-resolved sensors: (i) Predicting future EVE from present EVE, and (ii) Transforming HMI observations to AIA observations. These generic models may be re-applied to a wide variety of other problems not discussed in this section, for instance predicting future AIA from current AIA, or predicting EVE from AIA.

We stress that our baselines are not intended to be the top-performing solutions, but rather a rubric that shows how well a simple data-driven approach would perform. This serves two functions: First, future model implementations that are more complex should out-perform these baselines in the metrics we propose or other such metrics (e.g., focusing only on flaring events); and secondly, the baselines provide context necessary to properly evaluate a future model's performance. For instance, while a more complex model may achieve a low error rate such as 5\%, if our baseline already achieved a similar score, the complexity of the new model may not be warranted.

\subsection{EVE-to-EVE Prediction}\label{sec:EVEPRed}

The goal of this task is to predict future EVE observations given current EVE observations at a future time ranging from a few hours up to a full solar rotation. In order to provide statistically sound benchmarks in light of strong solar variability, we calculate the average relative error over all predictions for look-forward times of various fixed sizes. This statistical approach informs to what extent we can predict overall future phenomena for a given look-forward time, as well as account for strong solar variability.

There are two main sources of solar variability for this timescale. In shorter time-scales, the main source of variability are flares, which increase the EUV radiative output of the Sun by several orders of magnitude in time-scales of minutes and hours. The second is solar rotation itself (27 days at the synodic Carrington rotation rate). Rotation modulates EUV irradiance because active regions (bright in the EUV) have lifespans of 14 to 55 days and can come in and out of view as the Sun rotates. This active region permanence induces strong temporal correlations at look-forward times greater than 27 days as illustrated by the periodicity in Figure \ref{fig:future_predictions}, as the Sun's ``same face" rotates into view. For model evaluation, we choose a total look-forward time of 29 days, a duration long enough to expose the irradiance periodicity. 

For our input data we use the MEGS-A lines listed in Table \ref{tab:EVEWave} with the exception of Fe XVI 361~\AA~ which is the most sparsely measured line with only $\sim$1\% of the average number of line measurements.

~ \\
\noindent {\bf Baselines:} For this problem we report 3 simple baselines: 
\begin{enumerate}
    \item {\it Persistence.} This model assumes that all future observations of the Sun will be identical to its current state. Thus, for any time jump, it predicts that the future EVE observation will be the same as the current EVE observation.
    \item {\it Constant.} This model assumes that the Sun produces a constant EUV irradiance and therefore gives a constant prediction irrespective of the current EVE observation. We set this constant to the training set average per line. 
    \item {\it Linear.} This model assumes that future observations are a linear transformation of the current observations plus a constant bias. We fit this model using ridge regression, or a linear model with Tikhonov/L2 regularization. In particular, for a given spectral line and look-forward time, if $\xB_i$ is the current measurement and $y_i$ the corresponding future average observation, we solve for $\wB$ such that $\lambda ||\wB||^2_2 + \sum_i^n ||\wB^T \xB_i - y_i||$ is minimized for all instances $i$. We set $\lambda$ per model by grid search to minimize the average normalized absolute error, doing 2-fold cross validation on the training set.
\end{enumerate}

~\\
\noindent {\bf Results:} We evaluate the average normalized absolute error for these models for look-forward times ranging from 2 hours to 29 days in steps of 2 hours and report our results in Figure \ref{fig:future_predictions}. The linear and persistence both show trends corresponding to the solar rotation: their errors peak at approximately half a solar rotation and reduce steadily until a full rotation occurs, thus confirming the strong correlation between observations separated by exact rotations. The average model's error on the other hand is effectively constant; small variations occur because pairs of 1 day jump observations exist from January 1, 2012 up to December 30, 2013; while pairs of 29 day jumps can only be tested up to December 2, 2013.

Collectively, the results underscore the importance of having good baselines via the surprising effectiveness of even trivial models such as the persistence or average models. For instance, although the average model entirely ignores the current EVE observation, it is able to predict 7 of the 14 EVE lines with less than 10\% average normalized absolute error; and similarly, at a look-forward time of 27 days, 10 lines can still be predicted within 10\% error by the persistence model. 

It is true that the linear model frequently improves on the persistence model, especially for high-error lines like Fe XVI and Fe XV, and look-forward times much less than a full rotation. However, for many look-forward times and lines, the trivial persistence model actually outperforms the relatively complex linear model, demonstrating how simple baselines may assist in properly assessing the effectiveness of a machine learning model. 

\begin{table*}[t]
    \caption{Results for HMI to AIA prediction. The top performing method is underlined. \label{tab:hmitoaia}}
    \centering
    \begin{tabular}{
    @{}l@{~~}c@{~~}c@{~~}c@{~~}c@{~~}c@{~~}c@{~~}c@{~~}c@{~~}c@{~~}c@{~~~~~}c@{~~}c@{~~}c@{~~}c@{~~}c@{~~}c@{~~}c@{~~}c@{~~}c@{~~}c@{}} \toprule
            &  \multicolumn{10}{c}{Mean (Lower Better)} & \multicolumn{10}{c}{\% Pixels $<$ 10\% Error (Higher Better)} \\
            & Avg & 94 & 131 & 171 & 193 & 211 & 304 & 335& 1600& 1700 
            & Avg & 94 & 131 & 171 & 193 & 211 & 304 & 335& 1600& 1700 
    \\ \midrule
    3 Layer & 2.08 & 0.80& 0.98& 4.63& 5.39& 3.70& 0.85& 0.72& 0.73& 0.90& 15.4 & 12.0& 13.8& 12.1& 14.7& 12.5& 15.8& 11.1& 22.9& 23.5
    \\
    7 Layer & 0.83 & \underline{0.35}& \underline{0.38} & \underline{1.55}& 1.94& 1.27& 0.54& 0.45& 0.46& 0.52& 18.1 & 19.5& 17.7& 14.3& 15.6& 14.2& 15.2& 14.2& 24.3& 27.7
    \\
    11 Layer & \underline{0.75} & 0.37& 0.40 & 1.66& \underline{1.55} & \underline{1.06} & \underline{0.47} & \underline{0.37} & \underline{0.42} & \underline{0.47} & \underline{20.8} & \underline{20.3} & \underline{20.6} & \underline{16.4} & \underline{17.6} & \underline{16.3} & \underline{18.2} & \underline{17.4} & \underline{28.9} & \underline{31.4}
    \\ 
    \midrule
    &  \multicolumn{10}{c}{\% Pixels $<$ 20\% Error} & \multicolumn{10}{c}{\% Pixels $<$ 50\% Error (Higher Better)} \\
    3 Layer & 29.2 & 23.7& 27.0& 23.6& 28.8& 24.9& 31.1& 22.2& 40.5& 40.7& 58.0 & 53.9& 56.3& 49.8& 61.6& 57.4& 65.3& 54.4& 62.5& 60.6
    \\
    7 Layer & 34.5 & 38.3& 34.7& 27.7& 30.4& 27.9& 30.0& 28.2& 44.1& 49.5& 68.9 & \underline{80.0} & 74.5& 58.0& 64.0& 61.5& 65.1& 64.2& 74.5& 78.7
    \\
    11 Layer & \underline{39.5} & \underline{39.2} & \underline{39.6}& \underline{31.9} & \underline{34.3} & \underline{31.9} & \underline{35.5} & \underline{34.6} & \underline{52.2} & \underline{56.6} & \underline{75.0} & 77.5& \underline{76.4} & \underline{65.1} & \underline{70.6} & \underline{68.2} & \underline{73.8} & \underline{77.2}& \underline{81.6} & \underline{84.3}
    \\ 
    \bottomrule
    \end{tabular}
\end{table*}

\begin{figure*}
    \centering
    \includegraphics[width=\linewidth]{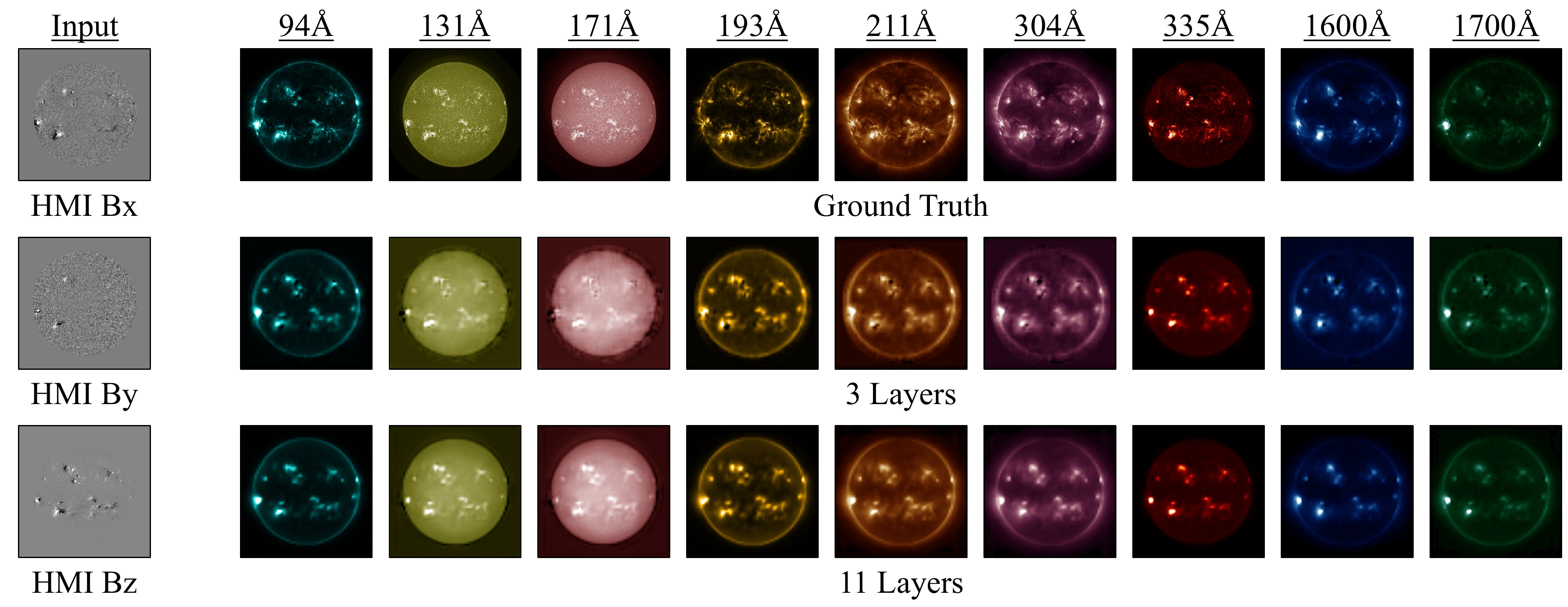}
    \caption{Results for HMI to AIA translation. The Left Panel shows the HMI inputs, while the Right Panel shows the ground-truth AIA (Top Panel) as well as the predicted AIA from a 3-layer network (Middle Panel) and 11-layer network (Bottom). While the 3-layer network performs well, additional layers (i) reduce artifacts (especially in 131~\AA and 171~\AA) presumably due to the depth, and (ii) better resolve coronal predictions (especially in 211~\AA and 304~\AA), presumably due to the larger receptive field caused by the additional layers.}
    \label{fig:hmitoaia}
\end{figure*}
 
\subsection{HMI-to-AIA Prediction}

We now move on to an example which demonstrates how a convolutional deep learning model may exploit the spatial richness of our dataset. In this application we show how a mapping between the HMI and AIA instruments is learned by treating it as an image-to-image translation problem. Such an approach is common in computer vision research, with applications as diverse as labeling each pixel in the scene with a category label (e.g., building) \cite{long_shelhamer_fcn}, generating images from sketches \cite{pix2pix2016}, inferring 3D properties of scenes \cite{wang15}, or detecting the pose of humans \cite{Cao17}.

We have physical reason to expect that there exists a mapping between the HMI and AIA instruments. While the HMI instrument infers information about the solar magnetic field from the solar photosphere, the AIA instrument measures UV/EUV emission from the solar chromosphere and corona. Since the chromosphere and corona are spatially structured by the presence of strong magnetic fields, UV/EUV emission will typically reflect information about the magnetic field through its spatial distribution, and \textit{vice versa}. Here we show how a simple convolutional model can realize the mapping from HMI to AIA.

~\\
\noindent {\bf Baseline:} Our baseline is a deep convolutional neural network. This is a function composed of alternating convolutions and non-linearities that maps one multi-channel image (e.g., a 3-channel 256x256 image) to another (e.g., a 9-channel 256x256 image). This function can be fit to a dataset of inputs (i.e., HMI) and desired outputs (i.e., AIA) via standard optimization procedures. Throughout we work with 256x256 images.

We adopt a basic approach for our network consisting of a three parts: (i) an initial feature extraction following ResNet \cite{he16resnet} consisting of a 7x7 convolution with stride 2 followed by 3x3 max-pooling with stride 2, which expands the receptive fields of subsequent feature maps; (ii) a variable number of 3x3 convolutions with stride 1; (iii) 3x3 convolution yielding a 9-channel prediction followed by 4x bilinear upsampling. All intermediate convolutions have 128 filters and are followed by a Rectified Linear Unit \cite{Nair10} and Batch normalization \cite{Ioffe15}. By varying the number of intermediate convolutions blocks in part ii, we can control both parameter count and effective receptive field of the network. We report results with 3, 7, and 11 layers (i.e., with 2, 6, and 10 hidden layers, including the initial convolution in part i).

We train the parameters of the network (e.g., filter weights and biases) via backpropagation and mini-batch stochastic gradient descent (SGD) to minimize the mean-squared error of the prediction. Specifically, we use SGD with Nesterov momentum \cite{Sutskever13} with momentum 0.99, weight decay $10^{-8}$, and batch size 32. We start with a learning rate of $10^{-3}$, which we multiply by $0.1$ every $5$ epochs, and train for 15 epochs. We checkpoint the network at the end of every epoch and take the network with lowest validation loss. To help learning, we divide inputs and outputs per-channel by their average over the training set (i.e., network is trained to predict the AIA 94~\AA~image divided by the empirical mean of AIA 94~\AA~over the training set). 

~ \\
\noindent {\bf Results:} We show sample qualitative results in Figure~\ref{fig:hmitoaia} for 3 and 11 layers. Even with a small number of hidden layers, a simple data-driven approach does a good job of getting the general shape and features of the sun, which suggests that results that get general features can be explained with relatively simple models and that more complex models must provide additional results compared to this. Adding more layers helps reduce artifacts at the edge of the disk in the 131~\AA~and 171~\AA~channels and more accurately resolve the corona in the 211~\AA~and 304~\AA~channels. The shallower network has difficulty accurately resolving the corona because each prediction is made from a small portion of the Sun; thus it produces a halo-like effect around the entire sun, as opposed to at specific locations on the disk. 

Quantitatively, increasing depth produces strong improvements, as seen in Table~\ref{tab:hmitoaia}. With a relatively unsophisticated deep network, $75\%$ of the pixels of AIA across all channels can be predicted within 50\% relative error from HMI observations.
As seen by the percentage good pixels metrics, 
1600\AA~and 1700\AA~observations appear to be among the easier to predict, and are almost always a few percentage points higher across both network depths and good pixel thresholds. This serves as a good sanity check on the results, since the photospheric and chromospheric brightness features in these two channels are known to be highly correlated to the photospheric distribution of magnetic fields.

\newpage
\section{Conclusion}\label{sec: conclusion}

In this paper we present a curated, high quality dataset from all three SDO instruments primed for machine learning research. We have preprocessed this data by downsampling AIA and HMI images from $4096 \times 4096$ to 512 x 512 pixels; removed \texttt{QUALITY} $\neq 0$ observations, corrected for instrumental degradation over time, and applied exposure corrections that account for Earth's elliptical orbit as well as AIA's automatic exposure control. We also have ensured that both AIA and HMI data are spatially co-located, have identical angular resolutions, and that all instruments are chrono-synchronous.

We also highlight some of the potential pitfalls of blindly applying machine learning techniques to solar data, or even more broadly:

\begin{enumerate}
  \item To maximize its versatility, SDO data products are nuanced and assume an expert-level understanding of its instrumental design and limitations.  Using them without this knowledge may lead to incorrect results and invalid conclusions.
  \item Most of the physical processes that drive solar variability occur at a much slower cadence than that of SDO's instruments (hours and days vs. minutes and seconds, respectively), requiring special care with the splitting of training, validation, and test sets. Splits must be performed along temporal blocks and not by random sampling, as is done in other settings with uncorrelated data samples. Random sampling in this case will lead to an overly optimistic estimate of validation error, leading to an inability to identify whether a model will generalize properly to future observations or has instead overfit to its data.
  \item Due to the relatively short timescales of solar variability, the simple forecasting models of permanence and climatological averages perform exceptionally well in hourly and daily timescales.  Due to this, error estimates of more advanced models are not meaningful in an absolute sense, but rather only when compared to these simple baselines.
\end{enumerate}

Finally, we provide a series of baselines that take advantage of this dataset to produce EVE time-forecasts and HMI$\rightarrow$AIA reconstructions. These examples are meant to illustrate some of the applications made possible by combining these data with ML techniques, as well as what heuristic performance measures one should expect to compare their own model implementations with.\\

As with many fields, Heliophysics has entered a data-rich age in which the human intellect alone is incapable of processing the copious amounts of data gathered by NASA's ever-growing spacecraft fleet.  Fortunately, the ongoing revolution in machine learning research will power a new age of data inference and physical insight that maximizes the scientific output of these data-rich missions.  It is important however for heliophysicists and computer scientists to work together to understand the properties and limitations of both the raw data and the ML techniques. If special care is not taken in understanding such limitations, we may unfortunately see a large amount of incorrect, overly optimistic, or worse, misleading research.  Interdisciplinary programs such as NASA's Frontier Development Laboratory can provide a vital common ground to facilitate this skill transfer and will be highly critical for the successful and fruitful development of ML techniques in the astrophysical sciences. 

\section*{Acknowledgements}
This project was conducted during the 2018 NASA Frontier Development Lab (FDL) program, a public/private partnership between NASA and SETI and industry partners including NVIDIA Corporation, Lockheed Martin and IBM. The authors thank IBM (especially Naeem Altaf) for generously providing computing resources on the IBM Cloud. We gratefully thank our mentors for guidance and useful discussion, as well as the SETI Institute for their hospitality. RG acknowledges support from the Moore-Sloan Data Science Environment at New York University and thanks Rob Fergus for useful discussion. The authors acknowledge support from NASA's SDO/AIA contract (NNG04EA00C) to LMSAL. AIA is an instrument onboard the Solar Dynamics Observatory, a mission for NASA's Living With a Star program. MCMC acknowledges support from NASA's Heliophysics Grand Challenges Research grant (NNX14AI14G).   \software{PyTorch \citep{paszke2017automatic}, ~ ~SunPy \citep{2015CS&D....8a4009S}, ~~SolarSoft \citep{Freeland1998}}.

\appendix
\hspace*{100pt}The dataset is available through the Stanford Digital Repository at:
\vspace*{3pt}

\begin{tabular}{lll}
\hspace*{-25pt}\url{https://purl.stanford.edu/vk217bh4910} & \url{https://purl.stanford.edu/jc488jb7715} & \url{https://purl.stanford.edu/dc156hp0190} \\
\hspace*{-25pt}\url{https://purl.stanford.edu/km388vz4371} & \url{https://purl.stanford.edu/sr325xz9271} & \url{https://purl.stanford.edu/qw012qy2533} \\
\hspace*{-25pt}\url{https://purl.stanford.edu/vf806tr8954} & \url{https://purl.stanford.edu/kp222tm1554} & \url{https://purl.stanford.edu/nk828sc2920}
\end{tabular}

\bibliography{main}
\bibliographystyle{aasjournal}

\end{document}